# Blind Channel Equalization

Mohammad Havaei, Sanaz Moshirian, Soheil Ghadami

*Abstract*— **Future services demand high data rate and quality. Thus, it is necessary to define new and robust algorithms to equalize channels and reduce noise in communications. Nowadays, new equalization algorithms are being developed to optimize the channel bandwidth and reduce noise, namely, Blind Channel Equalization.**

**Conventional equalizations minimizing mean-square error generally require a training sequence accompanying the data sequence. In this study, the result of Least Mean Square (LMS) algorithm applied on two given communication channels is analyzed. Considering the fact that blind equalizers do not require pilot signals to recover the transmitted data, implementation of four types of Constant Modulus Algorithm (CMA) for blind equalization of the channels are shown [2], [4]. Finally, a comparison of the simulation results of LMS and CMA for the test channels is provided.**

*Index Terms*— **Blind Channel Equalization, LMS algorithm, Constant Modulus Algorithm**

## I. INTRODUCTION

To invert the FIR (Finite Impulse Response) filter representing the channel, a so called equalizer is needed to combat the distortive channel effects. The purpose of an equalizer is to reduce the ISI (Inter Symbol Interference) as much as possible. Equalizer equalizes the channel – the received signal would be seen like it passed a delta response, as seen in figure.1.

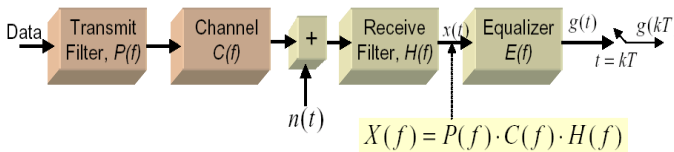

Figure.1.Communication System Model

Two different deconvolution problems:

- The system is unknown but its input is accessible and there for knowledge of the input signal is available. This is referred to as system identification. So a reliable model of the system is available and the requirement is to estimate the unknown input signal.

- The system in unknown and input is inaccessible this type of deconvolution is referred to as blind deconvolution.

Some assumptions are made in order to solve the blind deconvolution problem; if the input signal is a Gaussian process then information about the input is limited to second order statistics. For the solution to the blind deconvolution problem in the case of non minimum phase system to be feasible, the input must be non-Gaussian where higher order statistics than 2 are considered and non linear processing takes place [3].

Blind equalizers use properties in data format to recover the transmitted symbol. Non blind equalizers are data aided and use pilot sequences (training sequence) however since these results in occupation of the data package space, a reduction of system capacity occurs. A pre-known signal characteristic could be used to estimate the equalizer's coefficients. Using this technique, problems concerning sparse pilot sequences and fast changing channels are avoided [4], [2]. Equalizers which use signal symbols instead of pilot sequences are usually called blind equalizers.

## II. PROBLEM STATEMENT AND MAIN CONTRIBUTION

Many algorithms have been proposed during the past few decades however due to its simplicity, good performance and robustness the constant-modulus algorithm (CMA) is most often used in practice. Maybe the most prominent disadvantage of CMA is its slow convergence [3]. The traditional LMS algorithm also suffers from slow convergence as well. Although in practice the improved Normalized LMS (NLMS) is used, which usually converges many times faster than the LMS algorithm at the cost of only a few extra computations per update [4].

### A. Constant Modulus Algorithm

CMA is a stochastic gradient algorithm that minimizes the dispersion of the equalizer output around a circular contour. The CMA algorithm adapts filter coefficients at each time *n* in order to minimize the '2-2 modulus error', $\varepsilon^2$, [4]

$$\varepsilon^2 = \mathrm{E}\left[\left\||y(n)|^2 - A^2\right|^2\right] \qquad (1)$$

Where, *A* is the desired modulus, and

$$y(n) = \sum_{k=0}^{L-1}\omega_n(k)x(n-k) = W_n^t X_n \qquad (2)$$

Here *x(n)* is the signal to be equalized, *y(n)* is the complex equalizer output, $\omega_n(k)$ are the adaptive filter taps, and

$$X_n = [x(n)\ x(n-1)\ \cdots\ x(n-N+1)]^T$$
$$W_n = [\omega_n(0)\ \omega_n(1)\ \cdots\ \omega_n(N-1)]^T$$

The initial weight vector coefficients are set to zero. A stochastic gradient update of the filter coefficients is given by:

$$\omega_{n+1}(k) = \omega_n(k) - 4\mu(|y(n)|^2 - A^2)y(n)x^*(n-k) \\ = \omega_n(k) - \mu K x^*(n-k) \quad (3)$$

Where

$$K = 4(|y(n)|^2 - A^2)y(n) \quad (4)$$

$$W_{n+1} = W_n - 4\mu(|y(n)|^2 - A^2)y(n)X_n^* \\ = W_n - \mu K X_n^* \quad (5)$$

The step factor, $\mu$, is typically a constant selected through experimentation.

Of this family of algorithms, only the 1-1, 1-2, 2-1, and 2-2 forms are of much interest. As with the 2-2 CMA algorithm described above, the stochastic gradient updates take the general form:

$$W_{n+1} = W_n - \mu K_{p,q} X_n^* \quad (6)$$

Where

$$K_{1,1} = sign(|y_n| - A)\frac{y_n}{|y_n|} \quad (7)$$

$$K_{1,2} = 2(|y_n| - A)\frac{y_n}{|y_n|} \quad (8)$$

$$K_{2,1} = 2 sign(|y_n|^2 - A^2)y_n \quad (9)$$

### B. The LMS Algorithm

The LMS Algorithm is a developed form of steepest descent adaptive filter, in the family of stochastic gradient algorithms, which has a weight vector update equation given by:

$$\omega_{n+1} = \omega_n + \mu e(n) x^*(n) \quad (10)$$

Equation (10) is known as the LMS Algorithm. The simplicity of the algorithm comes from the fact that the update for the $k^{th}$ coefficient requires only one multiplication and one addition (the value for $\mu * e(n)$ need only be computed once and may be used for all of the coefficients). The update equation fot the $k^{th}$ coefficient is given by [1]:

$$\omega_{n+1}(k) = \omega_n(k) + \mu e(n) x^*(n-k) \quad (11)$$

The step size determines the algorithm convergence rate. Too small step size will make the algorithm take a lot of iterations while too large step-size will not convergence the weight taps. Step size may be calculated by rule of thumb as it is shown by (12) [1]:

$$\mu = \frac{1}{5(2N+1)P_R} \quad (12)$$

Where, N is the equalizer length, $P_R$ is the received power that could be estimated in the receiver.

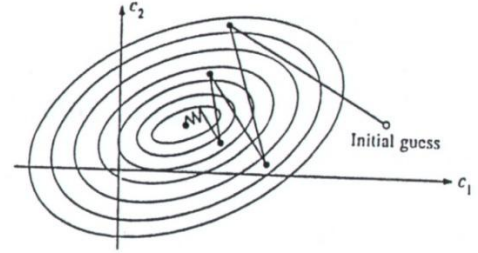

Figure.2. LMS convergence graph for the unknown channel of $2^{nd}$ order [1].

LMS algorithm convergence graph is illustrated in Figure.2. The initial weight vector coefficients are set to zero. The weights are updated each iteration by equation (10). This is basically going in the opposite of the gradient vector, until get the MMSE (Minimum Mean Square Error) is reached, meaning the MSE is 0 or a very close value to it. (In practice error of 0 cannot be reached since noise is a random process and the error could only be decreased to a value below the desired minimum).

### III. PROBLEM SOLUTION

In this project, two Gaussian communication channels with the following z-transform are considered [4]:

$$H_1 = \frac{1}{1 - 0.9Z^{-1}} \quad (13)$$

$$H_2 = \frac{1}{1 + 0.9Z^{-1}} \quad (14)$$

Moreover, it is known that the random transmitted signals are modulated by QPSK method and zero mean white Gaussian noise with a variance of $10^{-6}$ is added. If a signal passes through Channel $H_1$, the output for length-2 filters has little disparity eigenvalue, but it results in substantial eigenvalue disparity using channel $H_2$. In other word, channel 1 is more difficult than channel 2 to handle. Implementation of an equalizer of length 2 for CMA and length 8 for LMS is carried out. The step size for LMS is 0.007 and 0.001 for CMA. The simulation results for LMS algorithm, CMA(1,1), CMA(1,2), CMA(2,1), and CMA(2,2) for the first channel, $H_1$, are shown in Fig. 3, to Fig. 7 respectively. In these figures the top left, top right, bottom left and bottom right represent the transmitted signal, received signal, equalized signal and convergence plot:

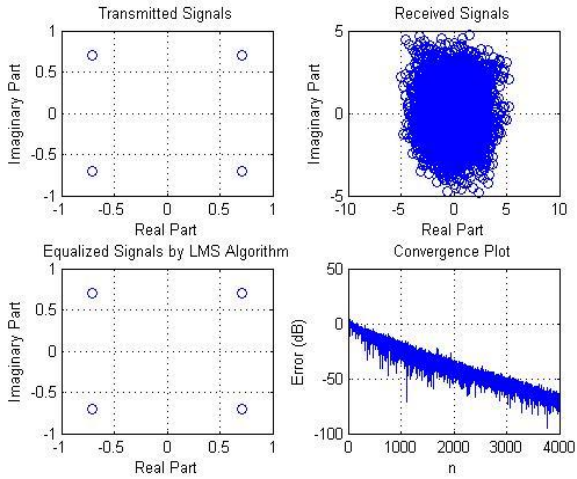

Fig. 3. Signal equalization based on LMS algorithm for channel 1.

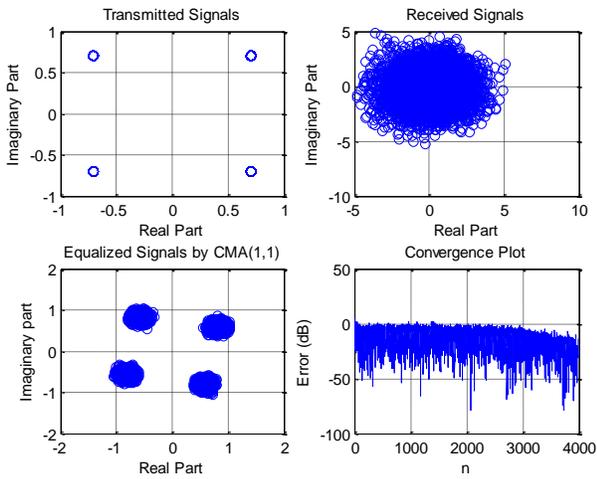

Fig. 4. Signal equalization based on CMA (1, 1) for channel 1.

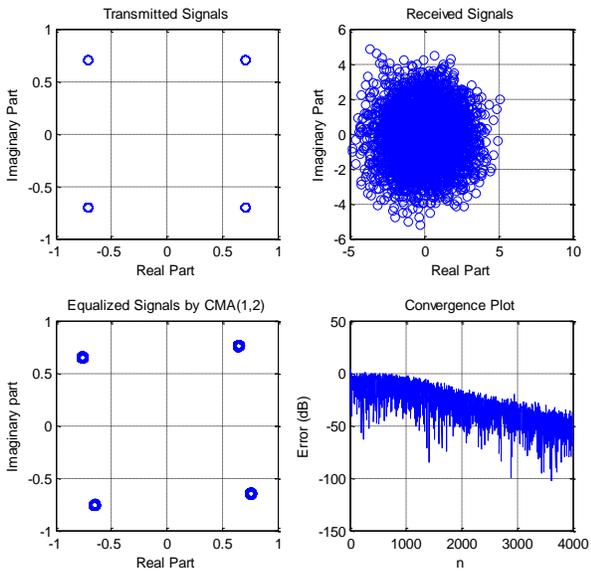

Fig. 5. Signal equalization based on CMA (1, 2) for channel 1.

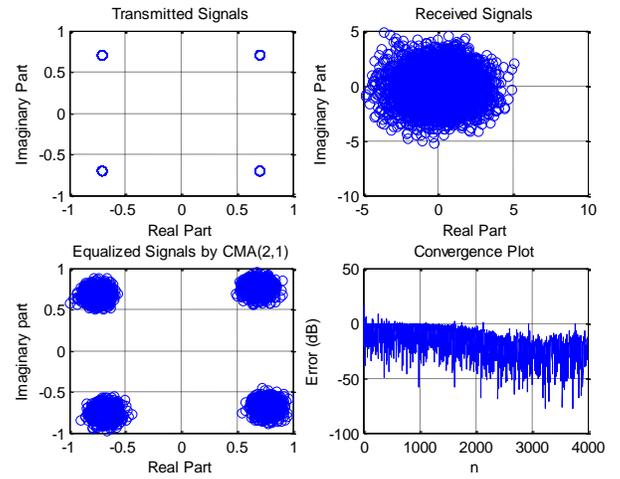

Fig. 6. Signal equalization based on CMA (2, 1) for channel 1.

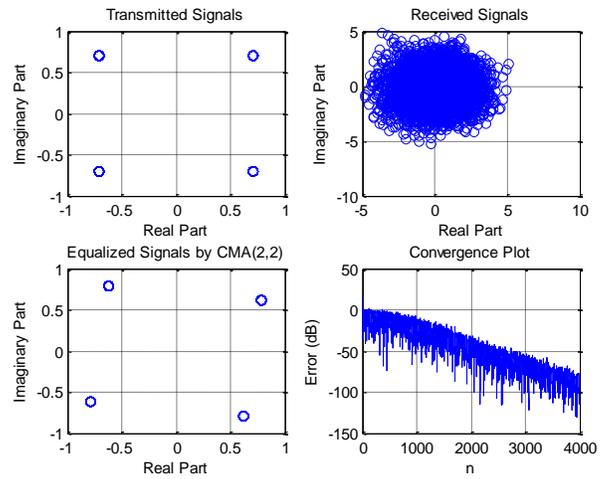

Fig. 7. Signal equalization based on CMA (2, 2) for channel 1.

Comparison between the convergence of LMS with each of CMA(1,1), CMA(1,2), CMA(2,1), and CMA(2,2) is done in Fig. 8, to Fig. 11 respectively:

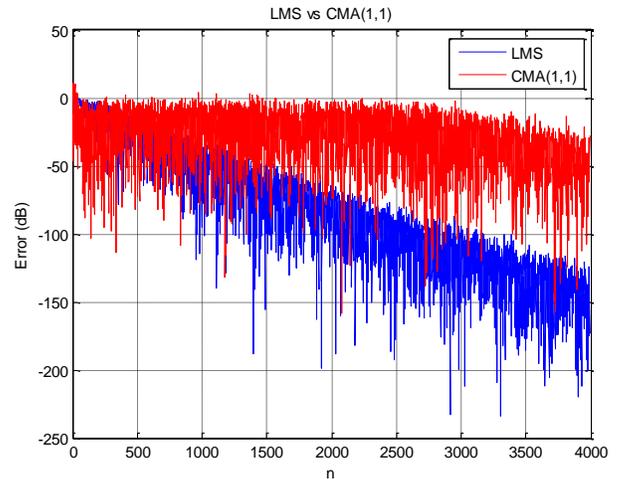

Fig. 8. Comparison of CMA (1, 1) and LMS for channel 1.

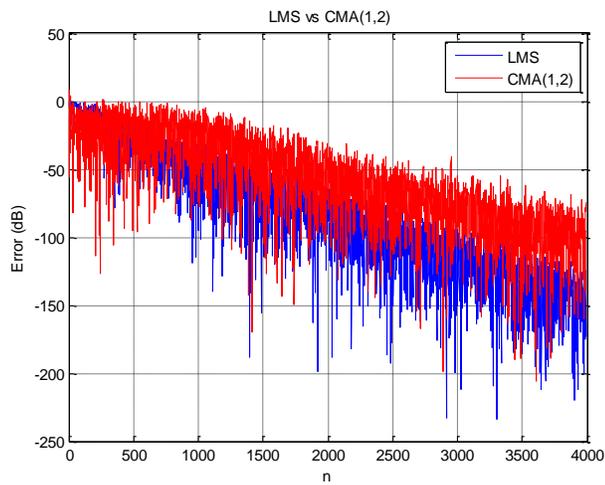

Fig. 9. Comparison of CMA (1, 2) and LMS for channel 1

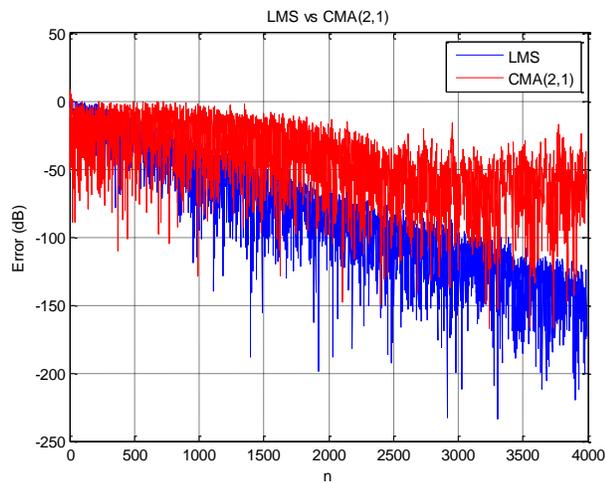

Fig. 10. Comparison of CMA (2, 1) and LMS for channel 1

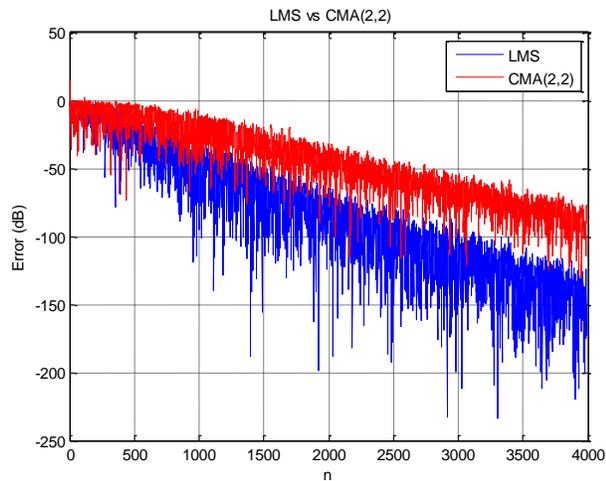

Fig. 11. Comparison of CMA (2, 2) and LMS for channel 1.

In the above figures, the blue curve represents the LMS algorithm while the red curve represents the CMA algorithms. As it is seen from the figures, LMS algorithm provides less error for channel 1. Among CMA algorithms, CMA (1, 2) has better results. Comparison of the least mean square error for different algorithms for channel 1 is also provided in Table I:

TABLE I
COMPARISON OF LEAST MEAN SQUARE ERROR OF DIFFERENT METHOD FOR CHANNEL 1

| Method | LMS | CMA(1,1) | CMA(1,2) | CAM(2,1) | CMA(2,2) |
|---|---|---|---|---|---|
| Error (dB) | -150 | -50 | -100 | -80 | -100 |

Similar experiment was carried out for the second channel. Simulation results for LMS algorithm, CMA (1,1), CMA(1,2), CMA(2,1), and CMA(2,2) for the second channel, $H_2$, are shown in Fig. 12, to Fig. 16 respectively:

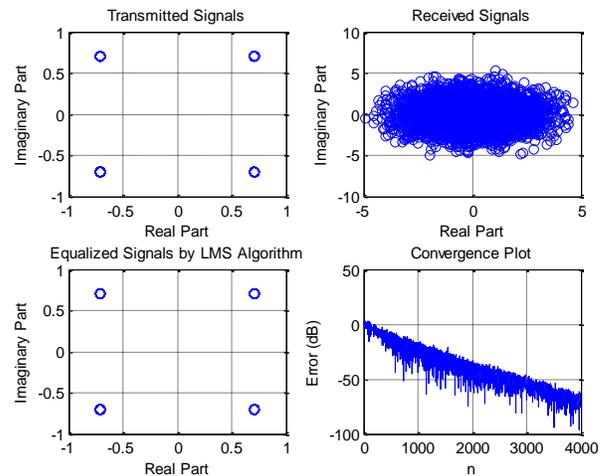

Fig. 12. Signal equalization based on LMS algorithm for channel 2.

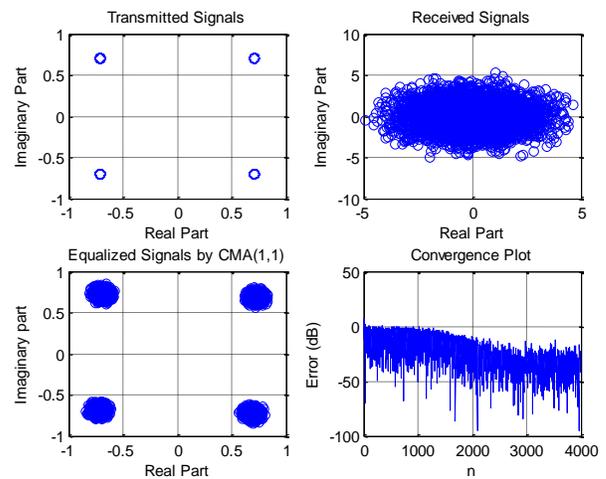

Fig. 13. Signal equalization based on CMA(1,1) algorithm for channel 2.

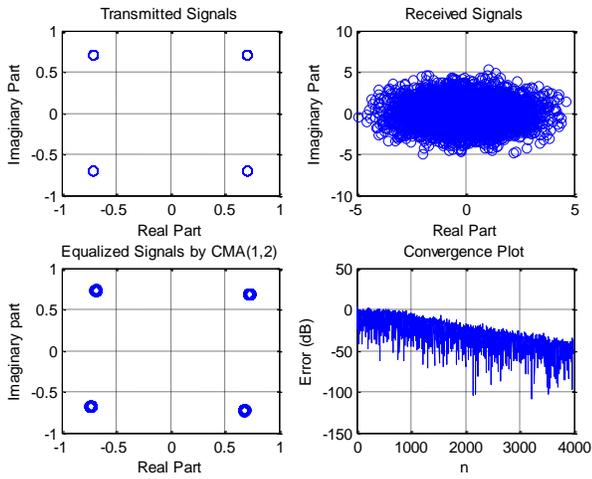

Fig. 14. Signal equalization based on CMA(1,2) algorithm for channel 2.

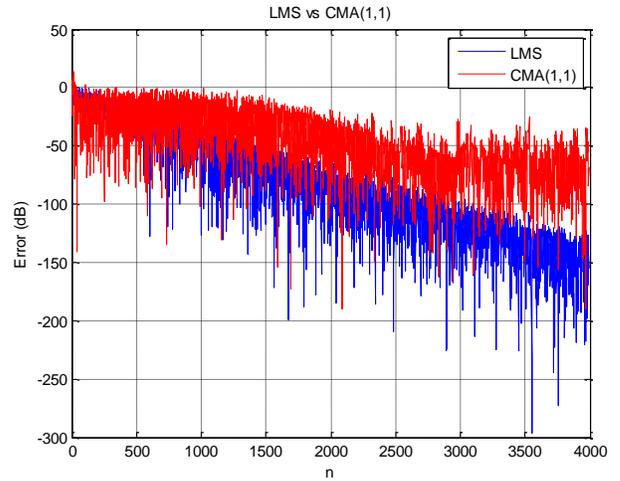

Fig. 17. Comparison of CMA (1,1) and LMS for channel 2.

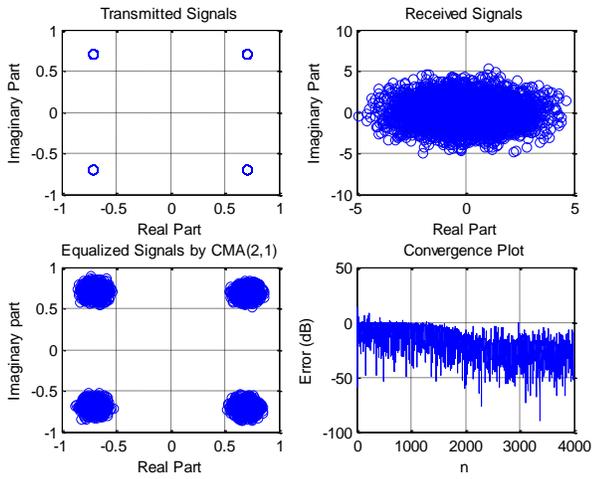

Fig. 15. Comparison of CMA (2,1) and LMS for channel 2.

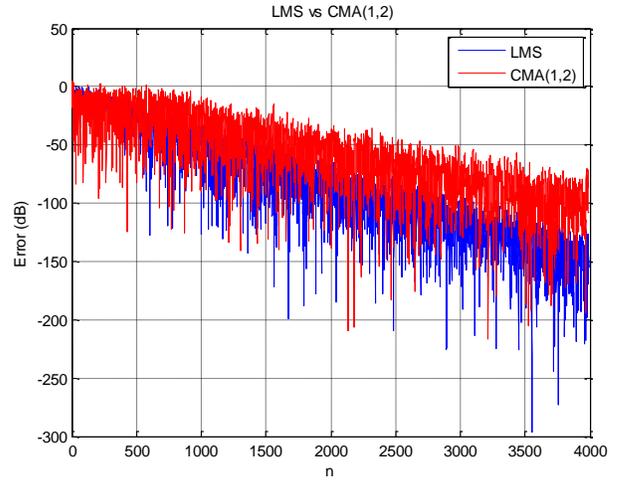

Fig. 18. Comparison of CMA (1,2) and LMS for channel 2.

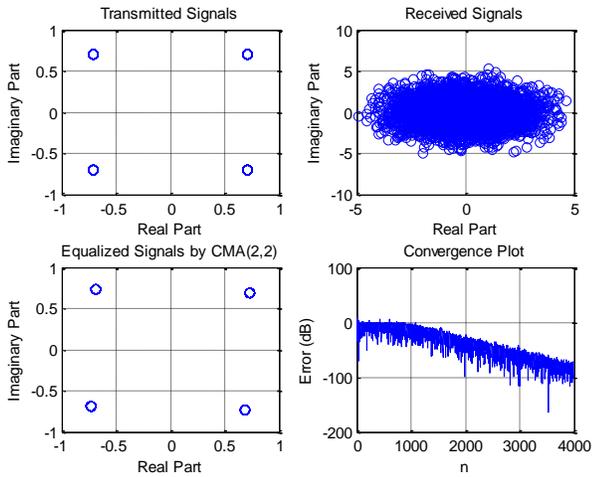

Fig. 16. Comparison of CMA (2,2) and LMS for channel 2.

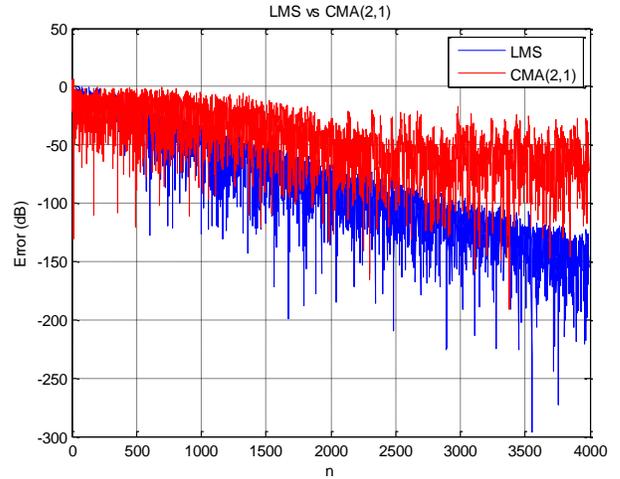

Fig. 19. Comparison of CMA (2,1) and LMS for channel 2

Comparison between convergence of LMS with that of each CMA(1,1), CMA(1,2), CMA(2,1), and CMA(2,2) is shown in Fig. 17, to Fig. 20 respectively :

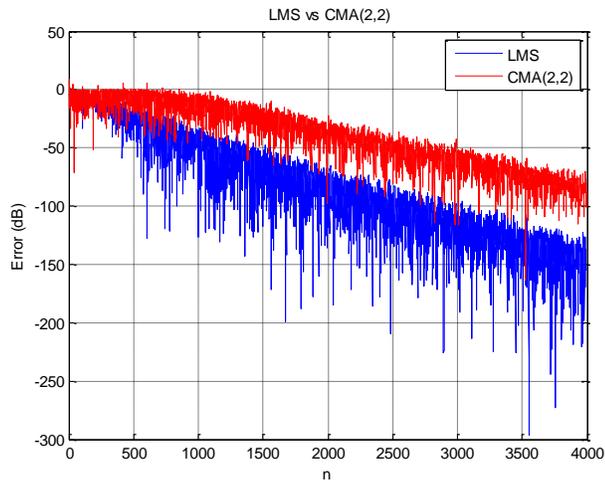
Fig. 20. Comparison of CMA (2, 2) and LMS for channel 2.

In these figures, the blue curve represents the LMS algorithm while the red curve represents the CMA algorithms. As it is seen in the figures, LMS algorithm provides less error for channel 2 in comparison with variants of CMA. Among CMA algorithms, CMA(1,2) has better results. Comparison of the least mean square error for different algorithms for channel 2 is provided in Table II:

TABLE II
COMPARISON OF LEAST MEAN SQUARE ERROR OF DIFFERENT METHODS FOR CHANNEL 2

| Method | LMS  | CMA(1,1) | CMA(1,2) | CMA(2,1) | CMA(2,2) |
|--------|------|----------|----------|----------|----------|
| Error  | -140 | -60      | -110     | -60      | -80      |

## IV. CONCLUSION

Blind channel equalization is used whenever there is no available training data, and we have no knowledge of coming data. From the results, it can be concluded that CMA(1,1), CMA(1,2), CMA(2,1), and CMA(2,2) perform slower than LMS algorithm, for channel 1 by the values of 100, 50, 70, 50 db and for channel 2 by the values of 70, 30, 70, 60 db respectively. It is also noted that initialization of the CMA algorithms has significant effect in the convergence rate.